\documentclass[amsmath,amssymb,aps,prl,reprint,superscriptaddress,footinbib,floatfix]{revtex4-1}
\def\apjs{Astrophys.~J.~Supp.~Ser.}
\def\apj{Astrophys.~J.}
\def\apjl{Astrophys.~J.~Lett.}

\def\mnras{MNRAS}
\def\prl{Phys.~Rev.~Lett.}
\def\pop{Phys.~Plasmas}

\usepackage{graphicx}
\usepackage{dcolumn}
\usepackage{bm}
\usepackage{amsmath,amssymb,framed,amsthm,xcolor} 
\usepackage{array,multirow,graphicx}
\usepackage{natbib}
\usepackage{float}

\usepackage{graphicx}
\usepackage{dcolumn}

\begin{document}
\title{Collisionless accretion onto black holes: dynamics and flares}
\author{Alisa Galishnikova}
\altaffiliation[alisag@princeton.edu]{}
\affiliation{Department of Astrophysical Sciences, Princeton University, 4 Ivy Lane, Princeton, NJ 08544, USA}

\author{Alexander Philippov}
\affiliation{Department of Physics, University of Maryland, College Park, MD 20742, USA}

\author{Eliot Quataert}
\affiliation{Department of Astrophysical Sciences, Princeton University, 4 Ivy Lane, Princeton, NJ 08544, USA}

\author{Fabio Bacchini}
\affiliation{Centre for mathematical Plasma Astrophysics, Department of Mathematics, KU Leuven, Celestijnenlaan 200B, B-3001 Leuven, Belgium}
\affiliation{Royal Belgian Institute for Space Aeronomy, Solar-Terrestrial Centre of Excellence, Ringlaan 3, 1180 Uccle, Belgium}

\author{Kyle Parfrey}
\affiliation{School of Mathematics, Trinity College Dublin, Dublin 2, Ireland}

\author{Bart Ripperda}
\altaffiliation[NASA Hubble Fellowship Program, Einstein Fellow]{}
\affiliation{Department of Astrophysical Sciences, Princeton University, 4 Ivy Lane, Princeton, NJ 08544, USA}\affiliation{School of Natural Sciences, Institute for Advanced Study, 1 Einstein Drive, Princeton, NJ 08540, USA}\affiliation{Center for Computational Astrophysics, Flatiron Institute, 162 Fifth Avenue, New York, NY 10010, USA}


\begin{abstract}
We study the accretion of collisionless plasma onto a rotating black hole from first principles using axisymmetric general-relativistic particle-in-cell simulations. We carry out a side-by-side comparison of these results to analogous general-relativistic magnetohydrodynamic simulations. Although there are many similarities in the overall flow dynamics, three key differences between the kinetic and fluid simulations are identified.  Magnetic reconnection is more efficient, and rapidly accelerates a nonthermal particle population, in our kinetic approach.   In addition, the plasma in the kinetic simulations develops significant departures from thermal equilibrium, including pressure anisotropy that excites kinetic-scale instabilities, and a large field-aligned heat flux near the horizon that approaches the free-streaming value. We discuss the implications of our results for modeling event-horizon scale observations of Sgr A* and M87 by GRAVITY and the Event Horizon Telescope.
\end{abstract}

\maketitle

{\em Introduction.---}The recent high-resolution images of synchrotron emission around the central black holes (BH) in M87 and the Milky Way (Sgr A*) captured by the Event Horizon Telescope (EHT) reveal asymmetric ring-like structures around the event horizon \cite{EHT1, EHT2022}. The radiation is produced by relativistic plasma on event-horizon scales. General-relativistic magnetohydrodynamic (GRMHD) simulations are a conventional tool for modeling accretion onto BHs \cite{EHTcode}. In conjunction with GR radiative transfer, one can predict aspects of the observed radiation, including spatially resolved images, from these numerical models \cite{GRRTcomparison}. This theoretical framework allows for a direct comparison of GRMHD simulations and observations. However, the accreting plasma in these systems is collisionless, which makes the simplifying assumptions of GRMHD formally inapplicable. Theoretical models thus require a kinetic approach, which describes collisionless plasmas from first-principles. In this Letter we present global GR kinetic simulations of BH accretion and determine the ways in which they differ from conventional fluid models.

Supermassive BHs show emission across the electromagnetic spectrum. Besides a relatively constant background emission, Sgr A* also exhibits episodic bright flares in the near-infrared and X-rays (e.g., \cite{Dodds-Eden_2009,Xray,GRAVITY}). The observed power-law emission implies a presence of accelerated particles (electrons and possibly positrons) near the BH. Studying the generation of non-thermal particles is not possible within GRMHD fluid models and requires a kinetic approach. Additionally, GRMHD does not accurately capture reconnection of magnetic field lines, which is conjectured to be responsible for particle energization and flares \cite{Markoff_2005, BroderickLoeb_2006, Dodds-Eden_2009, Dexter2020, Ripperda2020, Ripperda2022}. Specifically, the rate of reconnection, which regulates the energization efficiency and can be responsible for the duration of flares \cite{Bransgrove2021}, is known to be substantially faster in collisionless plasma (e.g., \cite{ Bhattacharjee2009,SironiSpitkovsky2014}). Ideal GRMHD also assumes an isotropic Maxwellian plasma distribution function, while collisionless plasmas easily develop pressure anisotropy along and across the magnetic field direction \cite{Kunz2014, Riquelme2015}, which leads to the development of plasma instabilities. The saturation of these instabilities regulates the thermodynamic state of the plasma, potentially affecting the global accretion dynamics and its observational properties.

\textit{Methods.} In order to study accretion of collisionless plasmas onto BHs, we perform  GR kinetic simulations using the particle-in-cell (PIC) code {\tt ZELTRON} which solves Maxwell's equations and the equations of motion for individual macroparticles in the 3+1 formalism \cite{GRPIC}. We also study the same problem with identical initial conditions using the GRMHD code {\tt Athena++} \cite{White_2016}, which allows for a side-by-side comparison of the two approaches. Both approaches utilize horizon-penetrating Kerr-Schild coordinates. We measure distance in units of the gravitational radius, $r_g = GM/c^2$, where $G$ is the gravitational constant, $M$ is the mass of the BH, and $c$ is the speed of light; time is measured in light crossing times of the gravitational radius, $r_g/c$.

In GRPIC, we resolve all the microphysical plasma scales and respect the correct hierarchy of scales, i.e. all plasma scales are significantly smaller than $r_g$. Due to their high computational cost, our simulations are limited to two-dimensional, axisymmetric accretion onto a BH in the $r-\theta$ plane, which is aligned with the BH spin, $a=0.95$. For simplicity, and motivated by the relevance to the accretion flow onto Sgr A* \cite{Quataert2003}, we start with a zero-angular-momentum spherically symmetric distribution of stationary plasma. Previous work shows that this accretion problem behaves similarly in many respects to one incorporating rotating initial conditions \cite{Ressler_2020}. We set an initially constant density, pressure, and magnetic field aligned with the spin axis throughout the box, and add randomly distributed magnetic loops \cite{Supplemental} to mimic magnetized turbulence expected in real accretion disks. 

We initialize a thermal plasma with $k_B T_{inj} \approx 0.02m_ic^2$, which corresponds to a Bondi radius $r_B= 2GM/c_s^2 \approx 50r_g$, where $k_B$ is the Boltzmann constant and $c_s$ is the sound speed. We focus on two initial values of the plasma-$\beta$ parameter, $\beta_0=P/P_{B_0}=4$ or $10$, where $P$ is the gas pressure and $P_{B_0}$ is the magnetic pressure of the initial vertical magnetic field $B_0$. Our GRPIC simulations use mass ratios of $m_i/m_e=1$ and $3$; the thermal Larmor radius of ions is set to be $\rho_L = v_{th} m_i c/eB_0 = 0.018 r_g$. Below we show results of simulations with mass ratio $m_i/m_e=1$, which we are able to run until $1000r_g/c$. The dynamics of ions in simulations with $m_i/m_e=3$ until $100r_g/c$ is similar compared to the case with $m_i/m_e=1$ \cite{Supplemental}.

Our simulation domain extends from the inner boundary, located below the event horizon, to the outer boundary at $100r_g$. We employ constant boundary conditions at the outer boundary in GRMHD, and use absorbing boundary conditions supplemented with injection of fresh plasma in GRPIC \cite{Supplemental}. In highly magnetized regions plasma density can get depleted; we therefore  apply a ceiling value for the magnetization parameter $\sigma=B^2/(4 \pi n (m_e+m_i) c^2) \approx 30$ in both approaches. In GRPIC simulations, we add electron-positron pairs when $\sigma$ is above this threshold, thus mimicking a pair cascade expected in these regions \cite{BeskinIstominParev1992, FFPairProduction1998, Crinquand2020, ChenYuan2020}.

\begin{figure}
    \centering
    \includegraphics[width=\columnwidth]{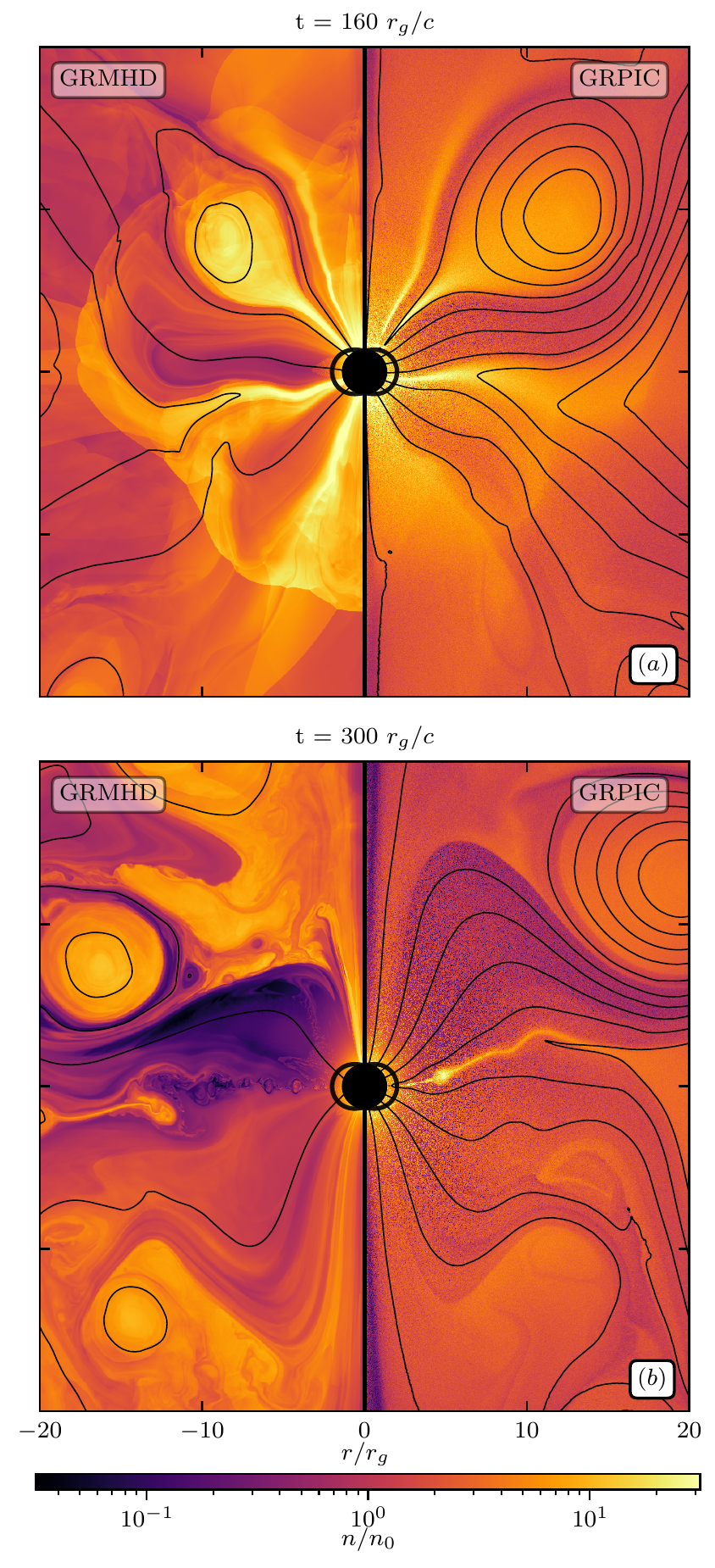}
    \caption{Comparison of fluid and kinetic simulations of accretion onto a rotating BH. The color shows $n/n_0$, which corresponds to the plasma number density in GRMHD (left) and ion and positron number density in GRPIC (right) in a region close to the BH ($20r_g$). The region inside the BH event horizon, $r_h=r_g(1+\sqrt{1-a^2})$, is shown by a black circle. Thin black lines represent the magnetic field lines, a thick black line outlines the ergosphere. The same quantities are shown at a time of $160r_g/c$ (a) and $300r_g/c$ (b). Several current sheets form and reconnect (a), leading to a flaring state (b).}
    \label{fig:2dcomparison}
\end{figure}

{\em Results.---}In Fig.~\ref{fig:2dcomparison} we show a side-by-side comparison of the evolution of the number density $n$ of the accreting plasma in GRMHD (plasma number density, left) and GRPIC (ion and positron number density, right) simulations with $\beta_0=4$ at a time of $160r_g/c$ (a) and $300r_g/c$ (b); $n_0$ corresponds to the initial value. Both panels show a zoom into the inner $20r_g$. Initially, the accreting plasma is free-falling onto the BH within the Bondi radius, dragging the magnetic field lines towards the event horizon (a). Inflow streams with a similar structure in both GRPIC and GRMHD are formed: a thin inflow is formed just below the equator, and a squeezed large loop is accreting above the equator. As the accretion proceeds, the BH's rotation and magnetic flux on the event horizon, which becomes dynamically important, lead to the launching of magnetically dominated outflows. The accretion stalls when the magnetic field becomes too strong, leading to thinning of the inflow streams into current sheets, onset of reconnection (b) \cite{Ripperda2020,Ripperda2022}, and the evacuation of the accretion flow in the equatorial plane (eruption). Eventually, the magnetic loop above the equator evacuates as an outflowing density bubble. Since the rate of collisionless reconnection is faster by a factor of a few, compared to MHD, the two numerical results ultimately diverge.

\begin{figure}[!ht]
    \centering
    \includegraphics[width=\columnwidth]{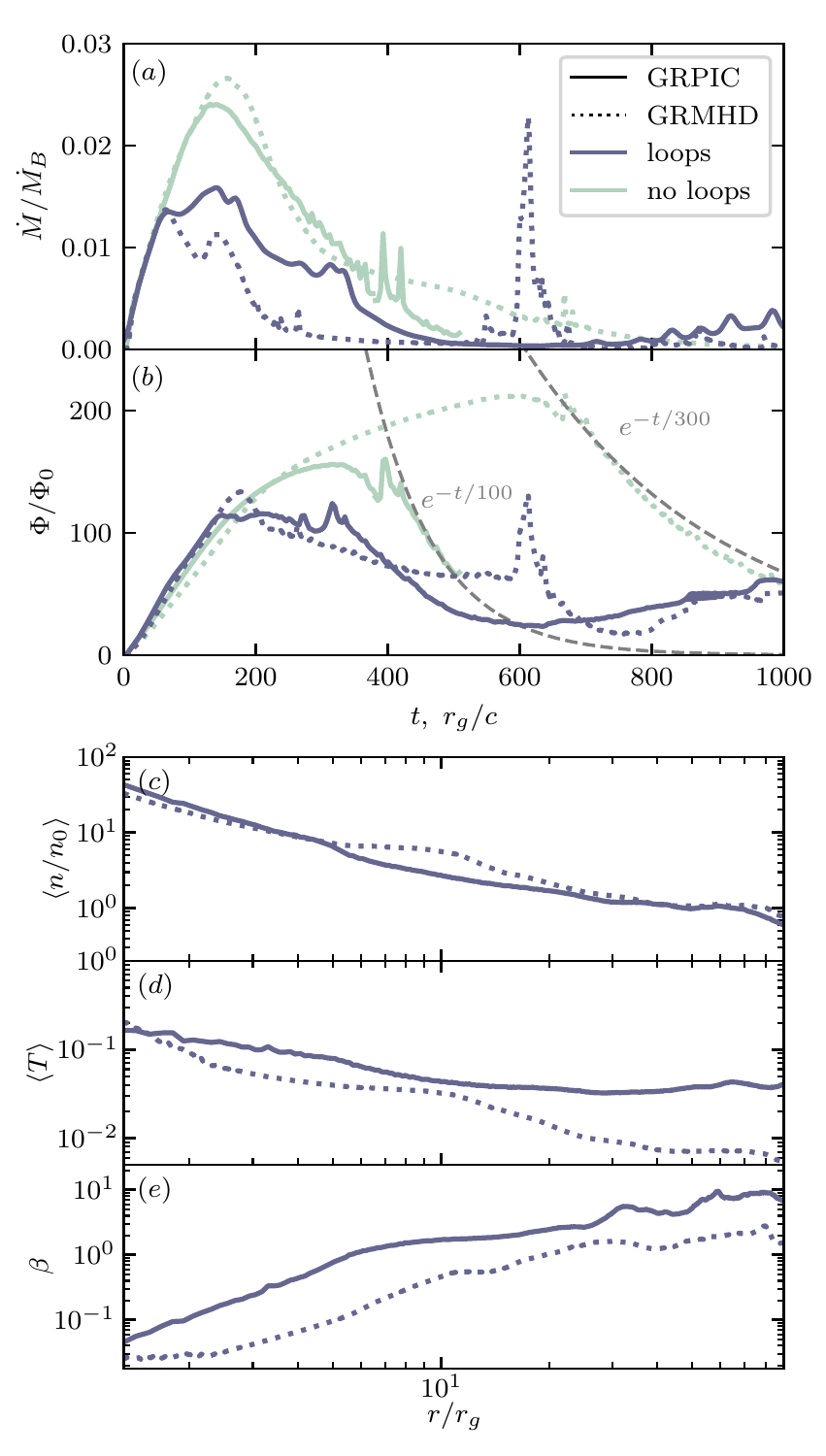}
    \caption{Evolution of averaged quantities in GRPIC (solid lines) and GRMHD (dotted). (a) Time evolution of the accretion rate in units of the Bondi accretion rate, $\dot{M}/\dot{M}_{B}$; (b) magnetic flux on the horizon normalized by its initial value $\Phi/\Phi_0$.  The gray dashed lines give the exponential fit for the decay rate of $\Phi/\Phi_0$, highlighting the difference of the reconnection rate. (c) Mean profiles of number density, $\langle n /n_0 \rangle$; (d) temperature, $\langle T \rangle$, and (e) plasma-$\beta = \langle P \rangle / \langle P_B \rangle$, as a function of radius $r/r_g$, averaged over $t=[100-200] r_g/c$. The runs initialized with the turbulent magnetic field are shown by a darker color (loops), with a vertical uniform magnetic field --- a lighter color (no loops).}
    \label{fig:time_evolution}
\end{figure}
The reconnection physics manifests itself in the time evolution of the accretion rate, $\dot{M}= - \int_{\theta} \int_{\phi} \sqrt{-g} \rho u^r d \theta d \phi$, and magnetic flux on the horizon, $\Phi=0.5 \int_{\theta} \int_{\phi} \sqrt{-g} \lvert B^r \rvert d \theta d \phi$, shown in Fig.~\ref{fig:time_evolution}a-b, which we normalize by the Bondi accretion rate, $\dot{M}_B$,  and initial value of the magnetic flux on the horizon, $\Phi_0$. Here, $g$ is the metric determinant, and $u^{\mu}$ is the fluid 4-velocity. Initially the infalling plasma in both approaches causes an increase in $\dot{M}/ \dot{M}_B$ and $\Phi/ \Phi_0$ at a similar rate, reaching saturation in the magnetically arrested state \cite{Tchekhovskoy2011}. GRMHD shows two $\dot{M}$ maxima followed by post-accretion eruption events associated with the onset of the decline of $\Phi$ at $\approx 200, 600r_g/c$. The GRPIC simulation, however, shows one $\dot{M}$ maximum followed by an eruption event at $\approx 380r_g/c$ and another accretion period starting at $\approx 800r_g/c$. Therefore, even though reconnection is more efficient, the variability --- the frequency of the eruption events --- might be smaller in the kinetic approach, as the accretion stalls due to its regulation by the efficient large-scale reconnection. Consequently, both $\dot{M}/\dot{M}_B$ and $\Phi/\Phi_0$ saturate at smaller values over a longer time period in GRPIC. 

A comparison of the two approaches due to the reconnection physics alone is demonstrated by simulations with an initially vertical uniform magnetic field (no loops, in Fig.~\ref{fig:time_evolution}a-b). Here, since kinetic reconnection is more efficient, GRPIC shows a steeper exponential decline in $\Phi/\Phi_0$ compared to GRMHD \cite{Bransgrove2021}. 

We show radial profiles (integrated over $\theta$ during the first accretion event, $100-200 r_g/c$) of density $\langle n/n_0 \rangle$ (c), temperature $\langle T \rangle$ (d), and $\beta=\langle P \rangle / \langle P_B \rangle $ (e) \cite{Supplemental}. We find a striking similarity of the number density profiles, while the temperature and $\beta$-profiles show a significant difference between the two approaches due to the non-ideal physics described next.
\begin{figure*}
    \centering
    \includegraphics[width=\textwidth]{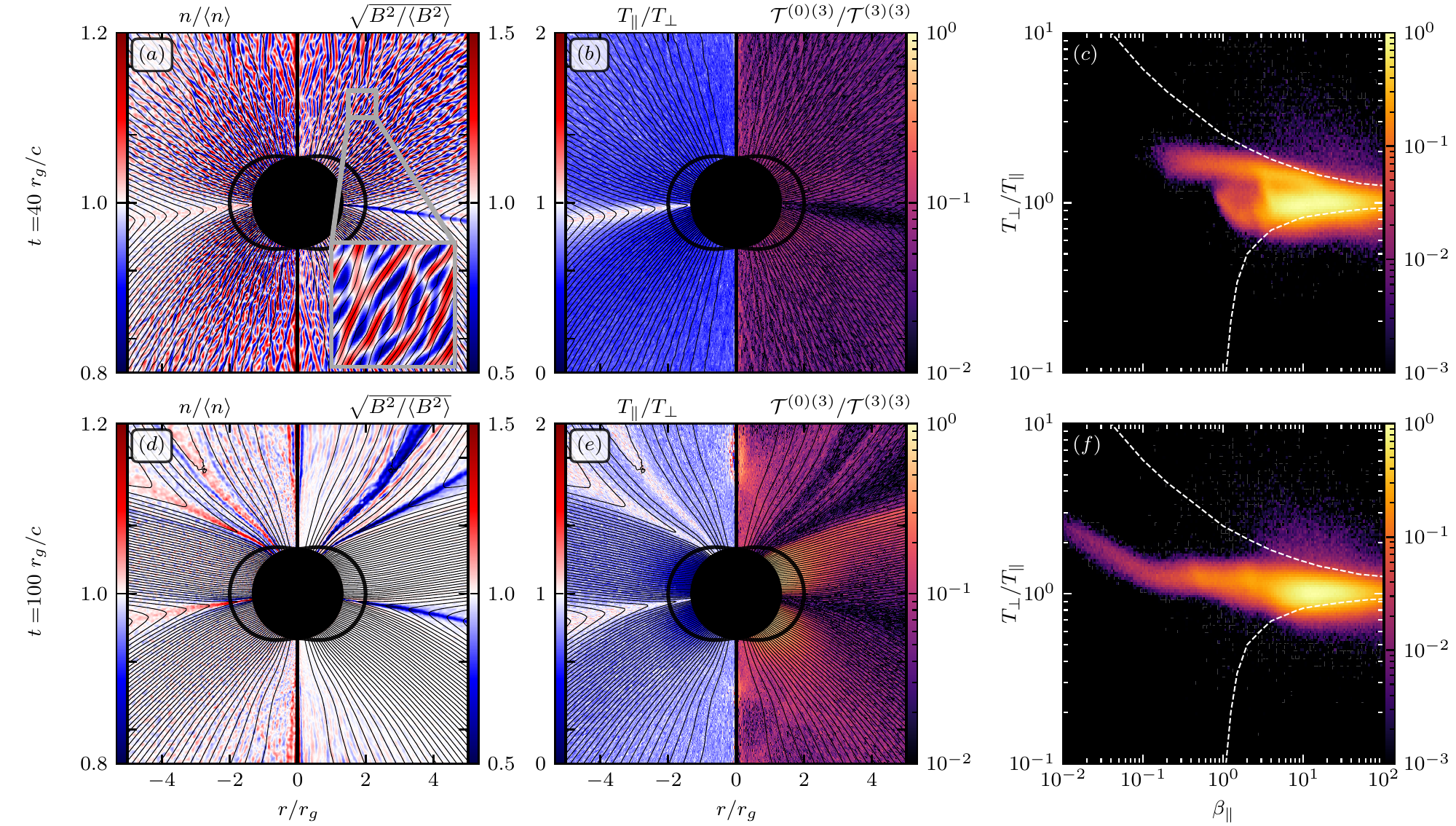}
    \caption{Anisotropy and heat flux in GRPIC simulation initialized with $\beta_0=10$. Each row corresponds to a different moment in time: $40r_g/c$ (a-c), and $100r_g/c$ (d-f). First column: number density (left) and magnetic field (right) fluctuations. Second column: temperature anisotropy $T_{\parallel}/T_{\perp}$ (left) and the ratio of non-ideal and ideal matter stress-energy tensor components in the tetrad frame $\mathcal{T}^{(0)(3)}/\mathcal{T}^{(3)(3)}$, which represents the ratio of parallel heat flux $q_{\parallel}$ and parallel pressure $P_{\parallel}$. The black circle and black lines represent the event horizon, ergosphere, and magnetic field lines as in Fig.\ref{fig:2dcomparison}. Third column: probability density as a function of $\beta_\parallel = P_{\parallel}/P_B$ and temperature anisotropy $T_{\perp}/T_{\parallel}$. White dashed lines correspond to the boundaries of the growth rate of mirror (top) and firehose (bottom) instabilities exceeding 10\% of the ion cyclotron frequency calculated using \cite{NHDS}. A mirror instability develops and saturates at $40r_g/c$, for which we show a zoom into the structure of the instability (a).}
    \label{fig:anisotropies}
\end{figure*}

To quantify another key feature of the kinetic approach, departure of the accreting plasma from thermal equilibrium, we calculate the matter stress-energy tensor in GRPIC, $\mathcal{T}^{\mu \nu}_{\rm matter}$, which can be used to derive the pressure tensor, $P^{\mu \nu}$, and the heat flux, $q^{\mu}$. We project these quantities onto a tetrad, $e^\mu_{(\nu)}$, where $e^\mu_{(0)}$ is directed along the (Eckart) fluid velocity and $e^\mu_{(3)}$ is along the magnetic field in the fluid frame. The pressure tensor in this tetrad frame is diagonal, $diag(P_\perp, P_\perp, P_\parallel)$, where parallel and perpendicular components are measured with respect to the magnetic field direction in the fluid frame. 

Ion quantities from a GRPIC simulation initialized with $\beta_0=10$ are shown in Fig.~\ref{fig:anisotropies}, where the two rows correspond to times of $40r_g/c$ and $100r_g/c$. The first column shows number density $n/ \langle n \rangle$ (left) and magnetic field $\sqrt{B^2/\langle B^2 \rangle}$ (right) fluctuations, second -- temperature anisotropy $T_{\parallel}/T_{\perp}$ (left) and ratio of the two stress-energy tensor components $\mathcal{T}^{(0)(3)}/\mathcal{T}^{(3)(3)}$ in the tetrad frame (right), where $\mathcal{T}^{(0)(3)}$ corresponds to parallel heat flux $q_{\parallel}$ (absent in ideal GRMHD). The last column shows an ion probability density plotted as a function of $\beta_\parallel = P_{\parallel}/P_B$ and $T_{\perp}/T_{\parallel}$. The polar inflow of the plasma leads to the build up of magnetic field, and an associated increase in $P_{\perp} \propto B$. The deviation from thermal equilibrium with $T_\perp > T_\parallel$ leads to the excitation of small-scale plasma density and magnetic field fluctuations (a), where we also show a zoom into a small region. This is a kinetic-scale mirror instability which develops when plasma crosses a $\beta$-dependent temperature anisotropy threshold, as shown in (c). The saturated strength of the magnetic field fluctuations, $|1-\sqrt{B^2/\langle B^2 \rangle}|$, is of order of a few tens of percent, consistent with local simulations \cite{Kunz2014, Riquelme2015, ICI}. At earlier times in the simulation, transiently, we observe the development of the electromagnetic firehose instability in the equatorial region, where $T_\parallel > T_\perp$ \cite{Supplemental}. 

At early times (b), the effective collisions due to particle scattering by the kinetic-scale fluctuations leads to a suppression of the heat flux, $\mathcal{T}^{(0)(3)}\approx 0.1\mathcal{T}^{(3)(3)}$. We find that  $q_\parallel$ is also $\approx 0.1$ of the value corresponding to the free streaming of particles along  magnetic field lines \cite{Supplemental}. As the accretion proceeds and the value of $\beta$ near the event horizon drops below $1$, the inflow is no longer accompanied by significant density or magnetic field fluctuations (d), which is consistent with the plasma being pushed away from the pressure-anisotropy instability boundaries (f). The absence of scattering on micro-scale magnetic field fluctuations leads to larger values of the non-ideal components of the stress-energy tensor, $\mathcal{T}^{(0)(3)}/\mathcal{T}^{(3)(3)}\approx 1$, outside of the current sheets (e), and $q_\parallel$ also approaches the free-streaming value. These non-ideal effects contribute to substantial differences between GRPIC and GRMHD temperature profiles (Fig.~\ref{fig:time_evolution}d). 

\begin{figure}[!ht]
    \centering
    \includegraphics[width=\columnwidth]{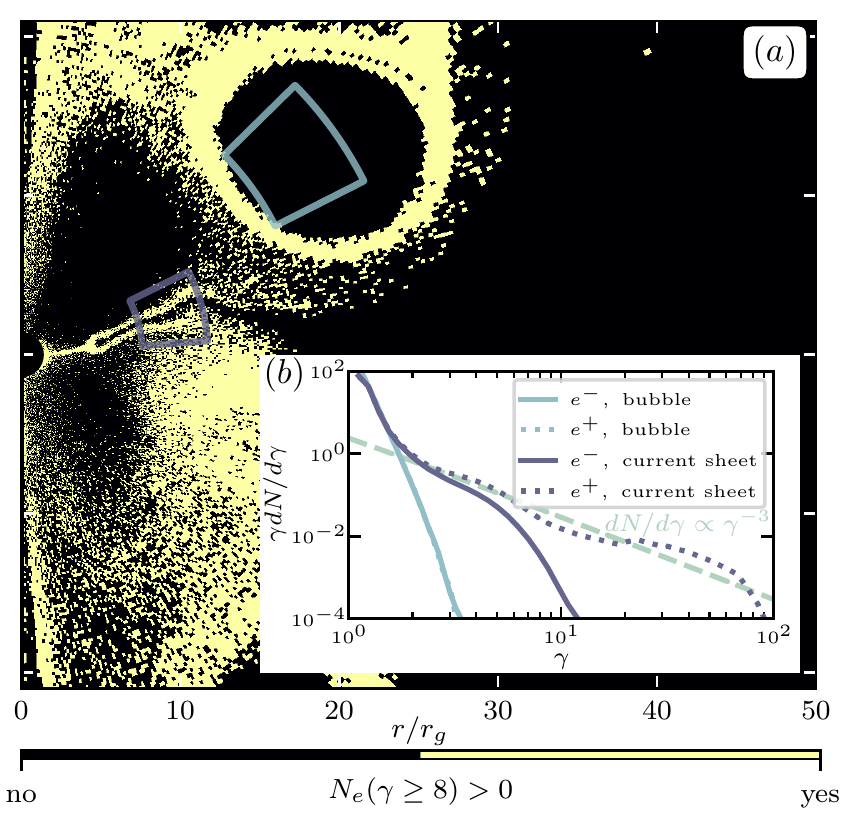}
    \caption{Panel (a) shows the presence (yes or no) of energetic electrons with $\gamma \geq 8$ throughout the GRPIC simulation at $300 r_g/c$ (shown in Fig.~\ref{fig:2dcomparison}b, right). To avoid low density regions near the axis, we show regions above the density threshold, $n>n_{0}/2$. Panel (b) shows particle spectra measured inside the outflowing bubble (blue) and in the current sheet (purple), which are outlined in (a) by the respective colors.}
    \label{fig:spectrums}
\end{figure}
To understand where particles are accelerated, we highlight regions with highly energetic particles with a Lorentz factor $\gamma \geq 8$ in GRPIC at $300 r_g/c$ in Fig.~\ref{fig:spectrums} (the same time snapshot as Fig.~\ref{fig:2dcomparison}b, right). These particles are predominantly located around the current sheets and the outflowing dense bubble. We measure particle spectra (b) in the regions outlined by corresponding colored wedges in (a). The spectral slope of $\approx -3$ (current sheet) is consistent with particle acceleration in relativistic magnetic reconnection for the measured magnetization parameter $\sigma\approx 5$ in the upstream \cite{SironiSpitkovsky2014,Guo2014,Werner2016}. As $\sigma$ increases due to evacuation of plasma in the jet region, we find a harder slope of $\approx -2$, consistent with higher $\sigma\gtrsim 10$ around the current sheet. Positively charged particles are accelerated more efficiently (Fig.~\ref{fig:spectrums}b) because their acceleration by the electric fields inside the current sheet is aligned with the direction of the outflow motion above it \cite{Cerutti2015}.  

{\em Discussion.---}A fully kinetic approach is crucial for understanding the dynamics of plasmas accreting onto supermassive BHs such as Sgr A* and M87*. Using global GRPIC simulations of accretion onto a rotating BH, we highlight three significant differences relative to matched GRMHD simulations: (1) differences in the physics of magnetic reconnection can lead to less frequent eruption episodes in GRPIC; (2) GRPIC includes pressure anisotropy with respect to the magnetic field and the associated kinetic instabilities; (3) in GRPIC a large field-aligned heat flux near the horizon is important in regulating the plasma temperature. Our kinetic approach allows for self-consistent modeling of particle acceleration during flaring episodes powered by magnetic reconnection, which opens up a unique opportunity for comparing theory with observed radiation spectra and light curves. Studying the relative heating and acceleration of ions and electrons will require a more realistic ion-to-electron mass ratio, which we currently lack in our simulations. In conjunction with general-relativistic radiative transfer, extension of our simulations to larger mass ratio and 3D will allow us to compare spatially resolved images, polarization maps and lightcurves constructed from GRPIC simulations to GRAVITY and EHT data. Future GRPIC simulations will rigorously measure the non-ideal corrections to the GRMHD stress tensor for a range of plasma conditions, which can then be included in GRMHD simulations \cite{Chandra_2015, Foucart2017} to improve their realism.

\begin{acknowledgments} 
This work was supported by NASA grant 80NSSC22K1054 and NSF grant PHY-2231698. EQ and AG were supported in part by a Simons Investigator grant from the Simons Foundation.  Computing resources were provided and supported by Princeton Institute for Computational Science and Engineering; and by the VSC (Flemish Supercomputer Center), funded by the Research Foundation Flanders (FWO) and the Flemish Government -- department EWI. This research is part of the Frontera computing project at the Texas Advanced Computing Center (LRAC-AST21006). Frontera is made possible by NSF award OAC-1818253. F.B. acknowledges support from the FED-tWIN programme (profile Prf-2020-004, project ``ENERGY'') issued by BELSPO. Support for this work was provided by NASA through the NASA Hubble Fellowship grant HST-HF2-51518.001-A awarded by the Space Telescope Science Institute, which is operated by the Association of Universities for Research in Astronomy, Incorporated, under NASA contract NAS5-26555. This research was facilitated by Multimessenger Plasma Physics Center (MPPC), NSF grant PHY-2206607. 
\end{acknowledgments}

\newpage
\widetext
\begin{center}
\textbf{\large Supplemental Materials}
\end{center}
\makeatletter

In this Supplemental Material we provide additional details about our fluid GRMHD and kinetic GRPIC simulations, as well as the calculation of the non-ideal stress-energy tensor in GRPIC.

\section{Setup}
We study accretion of plasma onto a spinning black hole with GRMHD and GRPIC simulations using boundary conditions described below. Both approaches utilize Kerr-Schild coordinates, which allows us to set the inner boundary inside the event horizon. At the outer boundary of the GRMHD simulations, we set constant boundary conditions which fix the fluid variables to their initial conditions. We mimic this in GRPIC simulations by keeping the plasma number density at a constant value of $n_e = 0.9 n_0$ at $r/r_g \in [95-96]$, close to the outer boundary, by injecting particles. We also introduce an absorption layer for the electromagnetic fields at the outer boundary, $r/r_g =100$, to allow an outflow of the escaping electromagnetic waves \citep{Cerutti2015}. 

MHD does not permit vacuum, hence we utilize radius-dependent density and pressure floors in GRMHD simulations, $\rho \geq 10^{-6} \rho_0 (r/r_g)^{-3/2}$ and $P \geq 3.33 \times 10^{-9} P_0 (r/r_g)^{-5/2}$, as well as additional limits for the plasma magnetization parameter, $\sigma = 2P_B/\rho \leq 30$, and $\beta=P/P_{B} \geq 10^{-3}$. In GRPIC, we utilize a pair-injection scheme close to the BH, $r < 15r_h$, by using a magnetization parameter threshold, $\sigma \approx 30$, above which we introduce electron-positron pairs into the system. Due to these floors, as the system evolves in time and $\beta$ decreases close to the BH, a substantial region becomes affected by the floors. Thus, the comparison of density and temperature profiles at late times has to be taken with caution. In more realistic three-dimensional case, accretion is not completely halted during the eruptions \citep{Tchekhovskoy2011}, and the effect of floors is less significant.

In both approaches, we initialize plasma at rest (all spatial components of $u^\mu$ are set to $0$) with a constant density ($\rho_0$ in GRMHD, $n_0$ in GRPIC) and temperature. The magnetic field is initialized via an out-of-plane vector potential, $A_{\phi}= A_0 + A_{\rm loops}$, represented by a sum of a uniform background vertical field along the $\hat{z}$-axis and additional randomly distributed magnetic loops:
\begin{subequations}
\begin{align}
    A_0 &= \frac{1}{2} B_0 r^2 \sin^2 \theta,\\
    A_n &= r (L_n-r_{c,n}) \exp{\Big(1 - \frac{L_n^2}{(L_n - r_{c,n}^2)} \Big)},\\
    A_{\rm loops} &= \begin{cases}
      0, \text{if} \  r_c < L_n\\
      \sum\limits_n^{N} k B_0 A_n, \ \text{otherwise},\\
    \end{cases}
\end{align}
\end{subequations}
Here, $B_0$ is the magnetic field strength in the absence of the loops, which corresponds to the initial plasma-$\beta_0$. Here $x_{c,n}$, $y_{c,n}$, and $L_n$ set the position of the $n$-th loop center and its size, $r_{c,n} = \sqrt{(x-x_{c,n})^2+(y-y_{c,n})^2}$ is the distance from the loop's center, $N=1000$ is the total number of loops. The normalization coefficient $k$ is chosen such that magnetic energy that is contained in the background vertical magnetic field is equal to the magnetic energy that is contained in the turbulent loops. Additionally, an exponential hole in $A_{\phi}$ and initial density is set by multiplying by an additional factor of $\exp{(5 (1 - 6/r) )}$ close to the BH at $r < 6r_g$, same as in \cite{Ressler_2020}. The loops are randomly distributed across the simulation box with random sizes ranging from $r_g$ to $20r_g$, avoiding the spin axis, $x_c> 20r_g$.

\begin{figure}[!ht]
    \centering
    \includegraphics[width=\textwidth]{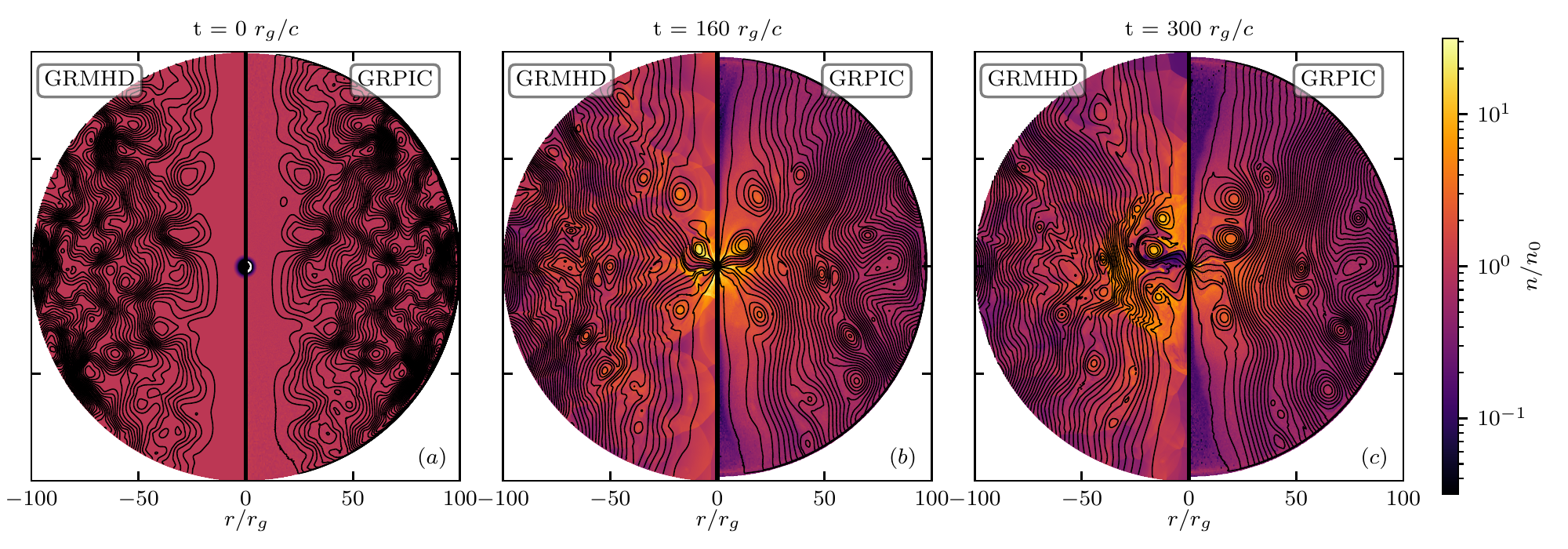}
    \caption{Comparison of fluid and kinetic simulations of plasma accreting onto a rotating BH. The color shows $n/n_0$, which corresponds to the plasma number density in GRMHD (left) and ion and positron number density in GRPIC (right) in a full simulation box (up to $100r_g$). Thin black lines represent magnetic field lines. The same simulations are shown at a time of $0 r_g/c$ (a), $160r_g/c$ (b), and $300r_g/c$ (c).}
    \label{fig:full_box}
\end{figure}
Our kinetic simulations are initialized with $20$ particles per cell ($10$ electrons and $10$ ions), which are thermally distributed with the thermal velocity, $v_{th}$, corresponding to the Bondi radius, $r_B \approx 50r_g$. In the absence of loops, the ion Larmor radius corresponds to $v_{th} m_i c /eB_0 \approx  0.018 r_g$. Our grid is logarithmically distributed in $r$, thus, the value of the electron skin depth $d_e=c/\sqrt{4 \pi n_0 e^2/m_e}$ at the outer cell sets the resolution of our simulations. 

We utilize horizon-penetrating Kerr-Schild coordinates, logarithmic in $r$. Our GRMHD simulations have resolution of $N_r \times N_\theta = 4096 \times 4096$ cells. GRPIC simulations with $\beta_0=4$ have a resolution of $N_r \times N_\theta = 7840 \times 7936$ cells, while those with $\beta_0=10$ have $N_r \times N_\theta = 11720 \times 11520$ cells, representing the largest-to-date global kinetic simulations. Figure \ref{fig:full_box}a shows the equivalent initial conditions ($t=0r_g/c$) for the GRMHD (left) and GRPIC (right) approaches with $\beta_0=4$. Color represents density, initially uniform with an exponential hole in the center, black lines represent magnetic field lines. We also show the time evolution of the setup at $t=160 r_g/c$ (b) and $t=300r_g/c$ (c) over the full box (compare to Fig.1 in the main text which shows the inner $20r_g$ of the computational domain).

\section{Calculation of the matter stress-energy tensor, pressure anisotropy and heat flux}
To quantify the departure of the accreting plasma from thermal equilibrium, we calculate the matter stress-energy tensor in the kinetic simulations for every species as $\mathcal{T}^{\mu \nu}_{{\rm matter}} \equiv\int \frac{d^3 p}{\sqrt{-g}p^t} p^{\mu}p^{\nu}f$, where $f$ is the distribution function, and $p^{\mu}$ is the particle contravariant momentum. The stress-energy tensor includes components that are present in ideal GRMHD and some that are not. To distinguish between these, we find the (Eckart) fluid velocity $u^{\mu} \equiv \frac {N^{\mu}}{\sqrt{N_{\nu} N^{\nu}}}$, where $N^{\mu} \equiv \int \frac{d^3 U}{\sqrt{-g}} (U^{\mu}/U^t) f$  and $U^{\mu}$ is the particle contravariant velocity. 
The pressure tensor is derived as the spatial part of the stress-energy tensor, $ P^{\mu \nu} \equiv \Delta^{\mu}_{\alpha}\Delta^{\nu}_{\beta} \mathcal{T}^{\alpha \beta}_{\rm matter}$, where $\Delta^{\mu \nu} \equiv g^{\mu \nu} + u^{\mu}u^{\nu}$ is the projection tensor. In a non-ideal plasma, the time-space components of the stress-energy tensor set the heat flux $q^{\mu} \equiv - \Delta^{\mu}_{\alpha} u_{\beta} \mathcal{T}^{\alpha \beta}_{\rm matter}$. The magnetic field in the fluid frame is expressed as $b^{\mu} = \frac{1}{2} \epsilon^{\mu \nu \kappa \lambda} u_{\nu} F_{\lambda \kappa}$, where $\epsilon$ is the Levi-Civita pseudo-tensor, and $F$ is the electromagnetic Maxwell tensor. The pressure tensor in a tetrad frame with $e^{\mu}_{(0)}=u^{\mu}$ and $e^\mu_{(3)}=\hat{e}_{b^\mu}$, is diagonal, with the perpendicular $P_{\perp}$ and parallel $P_{\parallel}$ fluid pressure components: $diag(P_\perp, P_\perp, P_\parallel)$. The pressure can then be used together with the number density $n$ to calculate the effective temperature: $k_B T_\parallel = P_\parallel/n$, $k_B T_\perp = P_\parallel/n$, and $T=(T_\parallel + 2T_\perp)/3$.

\begin{figure}[!ht]
    \centering
    \includegraphics[width=\textwidth]{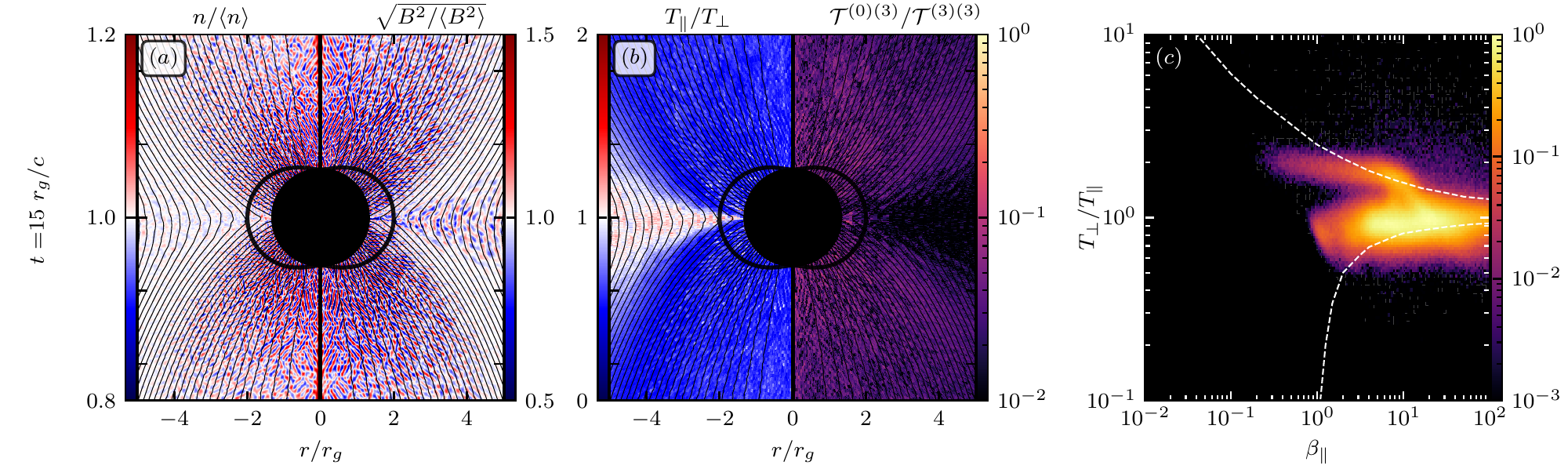}
    \caption{Anisotropy and heat flux for ions in GRPIC initialized with $\beta_0=10$ at $t=15r_g/c$. (a) Number density (left) and magnetic field (right) fluctuations. (b) Temperature anisotropy $T_{\parallel}/T_{\perp}$ (left) and the ratio of non-ideal and ideal matter stress-energy tensor components in the tetrad frame $\mathcal{T}^{(0)(3)}/\mathcal{T}^{(3)(3)}$, which represents the ratio of parallel heat flux $q_{\parallel}$ and parallel pressure $P_{\parallel}$. The inside of the event horizon is shown by a black circle in the center of the image, the thick black line outlines the ergosphere, and thin black lines represent the magnetic field lines. (c) Ion probability density as a function of $\beta_\parallel = P_{\parallel}/P_B$ and temperature anisotropy $T_{\perp}/T_{\parallel}$. White dashed lines correspond to thresholds for mirror (top) and firehose (bottom) instabilities. A mirror instability develops close to the BH; a firehose instability develops in the equatorial region.}
    \label{fig:anisotropies_s}
\end{figure}
\begin{figure}[!ht]
    \centering
    \includegraphics[width=0.47\textwidth]{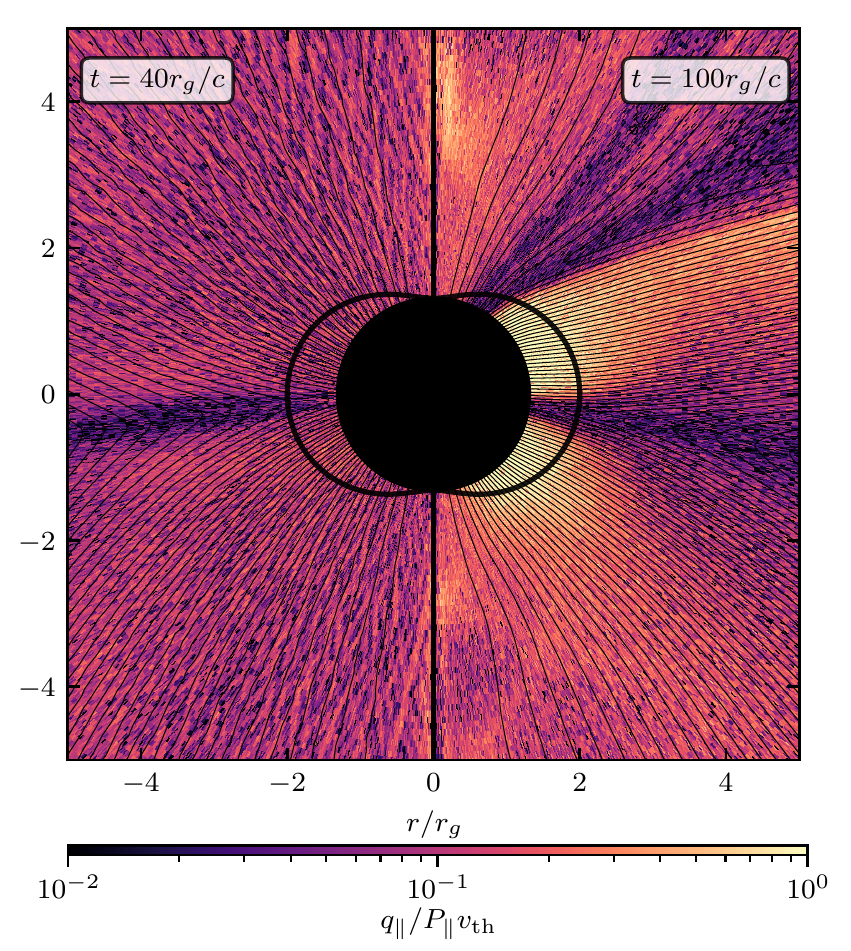}
    \caption{Parallel heat flux $q_{\parallel}$ as a fraction of the heat flux corresponding to particles free streaming along the magnetic field lines $P_{\parallel} v_{{\rm th}}$, for ions in GRPIC initialized with $\beta_0=10$. The Figure is a zoom-in of the region close to the BH ($r<5 r_g$). On the left a time of $40r_g/c$ is shown, on the right -- $100r_g/c$.}
    \label{fig:heat_flux}
\end{figure}
Figure ~\ref{fig:anisotropies_s} shows ion quantities from a GRPIC simulation initialized with $\beta_0=10$ but at an early time of $t=15r_g/c$, compare to Fig.3 in the main text which shows times of $t=40r_g/c$ and $t=100r_g/c$. We show number density $n/ \langle n \rangle$ (a, left) and magnetic field $\sqrt{B^2/\langle B^2 \rangle}$ (a, right) fluctuations, temperature anisotropy $T_{\parallel}/T_{\perp}$ (b, left), ratio of the two stress-energy tensor components $\mathcal{T}^{(0)(3)}/ \mathcal{T}^{(3)(3)}$ in the tetrad frame (b, right), and ion probability density as a function of $\beta_\parallel = P_{\parallel}/P_B$ and $T_{\perp}/T_{\parallel}$. White dashed lines in Fig.~\ref{fig:anisotropies_s}c correspond to the boundaries of the growth rate of mirror (top line) and firehose (bottom line) instabilities exceeding 10\% of the ion cyclotron frequency calculated using \cite{NHDS} for our case, $m_i/m_e=1$. Build-up of the magnetic field near the BH due to the polar inflow of the plasma leads to an associated increase in $P_{\perp} \propto B$. Thus, most of the plasma near the BH is dominated by the $T_{\parallel}/T_{\perp}<1$ region (b, left), and the mirror instability dominates. Because of our limited separation between macroscopic and microscopic scales, the plasma transiently significantly overshoots the mirror stability boundary \cite{Riquelme2015}, as seen in Figure ~\ref{fig:anisotropies_s}c. This effect is negligible for realistic accretion flows. Another anisotropy-driven electromagnetic instability develops at the equator (a, right), where $T_{\parallel}/T_{\perp}>1$ (b, left). Here the magnetic field fluctuations are not accompanied by plasma density fluctuations, which is a signature of a firehose instability. 

\begin{figure}[!ht]
    \centering
    \includegraphics[width=\textwidth]{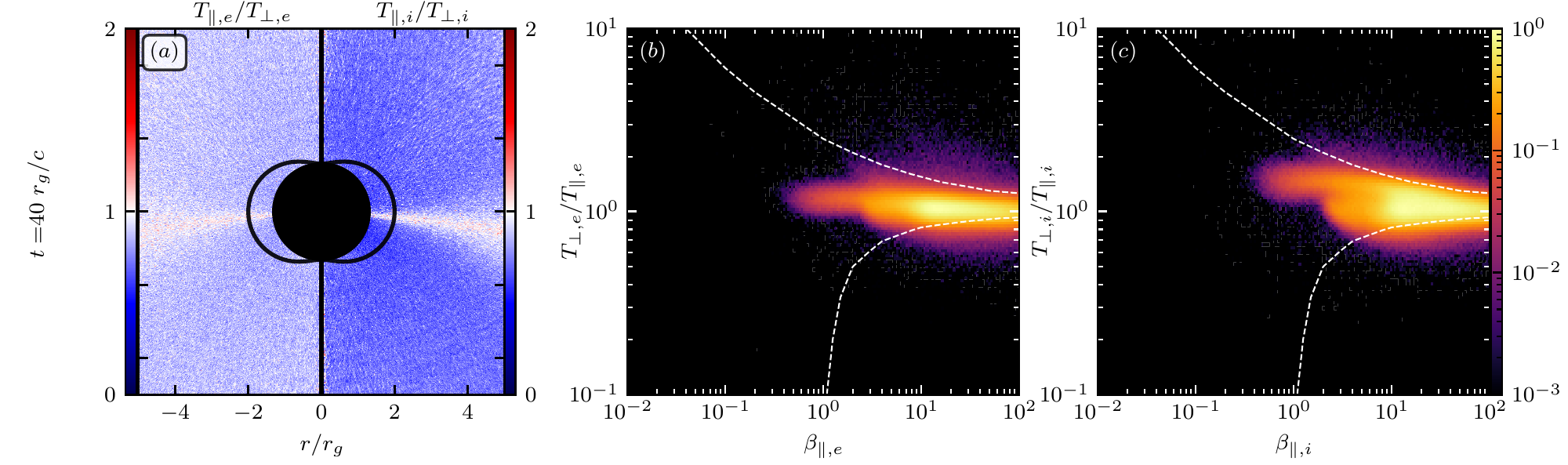}
    \caption{Anisotropy in GRPIC simulation with $m_i/m_e=3$ initialized with $\beta_0=10$ at $t=40r_g/c$. (a) Temperature anisotropy for ions $T_{\parallel,i}/T_{\perp,i}$ (left) and electrons $T_{\parallel,e}/T_{\perp,e}$ (right). Electron (b) and ion (c) probability density as a function of $\beta_\parallel = P_{\parallel}/P_B$ and temperature anisotropy $T_{\perp}/T_{\parallel}$. White dashed lines correspond to thresholds for mirror (top) and firehose (bottom) instabilities.}
    \label{fig:mime3}
\end{figure}

Figure \ref{fig:heat_flux} shows the ratio of the heat flux, $q_{\parallel}$, to the value corresponding to the free streaming of particles along magnetic field lines, $P_{\parallel} v_{{\rm th}}$, for ions in our GRPIC simulations with $\beta_0=10$, at two different times, $t=40r_g/c$ and $t=100r_g/c$. At the earlier time, the kinetic-scale mirror instability dominates in the region close to the BH, suppressing the heat flux, $q_{\parallel} \sim 0.1 P_{\parallel} v_{{\rm th}}$. At the later time, in the absence of effective scattering due to kinetic-scale instabilities, the heat flux approaches the free-streaming value in low-$\beta$ regions. Thus, we find that the heat flux is a significant fraction of the free-streaming value, even higher than what was found in previous non-ideal GRMHD simulations \citep{Foucart2017}.

Figure \ref{fig:mime3} demonstrates anisotropies measured at a time of $t=40r_g/c$ in a GRPIC simulation with $m_i/m_e=3$ initialized with $\beta_0=10$. We show temperature anisotropy for electrons $T_{\parallel,e}/T_{\perp,e}$ (a, left) and ions $T_{\parallel,i}/T_{\perp,i}$ (a, right), electron (b) and ion (c) probability density as a function of $\beta_\parallel = P_{\parallel}/P_B$ and $T_{\perp}/T_{\parallel}$.  White dashed lines in (b) and (c) correspond to the boundaries of the growth rate of mirror (top line) and firehose (bottom line) instabilities exceeding 10\% of the ion cyclotron frequency calculated using \cite{NHDS} for the case of $m_i/m_e=3$. Pressure anisotropy of ions in this simulation is similar to the one in the simulation with $m_i/m_e=1$ (Figure 3 in the main text). The probability density of ions (c) extends up to the mirror instability threshold. Scattering of electrons on the ion-induced fluctuations leads to their lower degree of the anisotropy. At larger values of the mass ratio, we expect excitation of electron-scale waves, which may lead to qualitatively different saturated values of the electron anisotropy.

\bibliographystyle{unsrt}

\end{document}